\documentclass[12pt]{article}
\usepackage{amsfonts}
\usepackage{amsmath}
\usepackage{amssymb}
\usepackage{graphicx}
\usepackage{color}
\usepackage[all, knot]{xy}
\usepackage{tikz}
\usepackage{array}

\usepackage{ulem}

\usepackage[utf8]{inputenc}
\usepackage{epstopdf}
\usepackage[footnotesize]{caption}
\usepackage{amsthm}
\usepackage{enumitem}
\usepackage{mathrsfs}

\usepackage[margin=3cm]{geometry}

\def \be {\begin{equation}}
\def \ee {\end{equation}}
\def \bea {\begin{eqnarray}}
\def \eea {\end{eqnarray}}
\def \nn {\nonumber}

\def \rr {\raise.35ex\hbox{\small $\prime$}\kern-.17em{\mbox{\large $\imath$}}}

\def \dels {\partial\kern-.6em /\kern.1em}
\def \As {{A\kern-.5em / \kern.5em}}
\def \Ds {D\kern-.7em / \kern.5em}

\def \ks {k\kern-.5em /}
\def \ls {l\kern-.5em /}

\newcommand{\ci}[1]{}

\newcommand{\ba}{\begin{eqnarray}}
\newcommand{\ea}{\end{eqnarray}}
\newcommand{\bal}{\begin{align}}
\newcommand{\eal}{\end{align}}
\newcommand{\bay}[1]{\left(\begin{array}{#1}}
\newcommand{\eay}{\end{array}\right)}

\newcommand{\hide}[1]{}

\newlist{axioms}{enumerate}{2}
\setlist[axioms,1]{label=\textbf{A\arabic{axiomsi}.}, ref=A\arabic{axiomsi}}
\setlist[axioms,2]{label=\textbf{A\arabic{axiomsi}\rlap{\myEnumCounter{axiomsii}}.},%
                   ref=A\arabic{axiomsi}\myEnumCounter{axiomsii},%
                   align=parleft,%
                   leftmargin=0em,%
                   itemsep=1.4ex,%
                   before={\stepcounter{axiomsi}}}

  \usetikzlibrary{decorations.markings}

\begin{document}
\begin{titlepage}

\begin{center}

\textbf{\LARGE
Accurate Study from\\
 Adaptive Perturbation Method
\vskip.3cm
}
\vskip .5in
{\large
Chen-Te Ma$^{a, b, c, d}$ \footnote{e-mail address: yefgst@gmail.com}
\\
\vskip 1mm
}
{\sl
$^a$
Guangdong Provincial Key Laboratory of Nuclear Science,\\
 Institute of Quantum Matter,
South China Normal University, Guangzhou 510006, Guangdong, China.
\\
$^b$
School of Physics and Telecommunication Engineering,\\ 
South China Normal University, Guangzhou 510006, Guangdong, China.
\\
$^c$
Guangdong-Hong Kong Joint Laboratory of Quantum Matter,\\
 Southern Nuclear Science Computing Center, 
South China Normal University,
 Guangzhou 510006, China.
\\
$^d$
The Laboratory for Quantum Gravity and Strings,\\
 Department of Mathematics and Applied Mathematics,\\
University of Cape Town, Private Bag, Rondebosch 7700, South Africa.
}
\\
\vskip 1mm
\vspace{40pt}
\end{center}
\begin{abstract}
The adaptive perturbation method decomposes a Hamiltonian by the diagonal elements and non-diagonal elements of the Fock state. 
The diagonal elements of the Fock state are solvable but can contain the information about coupling constants. 
We study the harmonic oscillator with the interacting potential, $\lambda_1x^4/6+\lambda_2x^6/120$, where $\lambda_1$ and $\lambda_2$ are coupling constants, and $x$ is the position operator. 
In this study, each perturbed term has an exact solution. 
We demonstrate the accurate study of the spectrum and $\langle x^2\rangle$ up to the next leading-order correction. 
In particular, we study a similar problem of Higgs field from the inverted mass term to demonstrate the possible non-trivial application of particle physics.  
\end{abstract}
\end{titlepage}

\section{Introduction}
\label{sec:1}
\noindent
{\it Perturbation method} is a known approximation for studying non-solvable systems \cite{Hioe:1978jj}. 
The known procedure is to begin from the non-interacting system and then do a perturbation from the coupling terms. 
Therefore, people cannot apply the procedure to the strongly coupled region. 
Applying the perturbation method to Quantum Field Theory (QFT) builds a generic tool for a probe of a weakly coupled region. 
Although people still do not know how to study a strongly coupled region from a similar procedure \cite{Ma:2018efs}, various physical phenomena and experiments were confirmed by the perturbation method. 
The most interesting problem in the strongly interacting system should be Quantum Chromodynamics (QCD). 
QCD describes the dynamics between quarks and gluons. 
The color confinement and asymptotic freedom are open questions, and it is necessary to extract physical information from the low-energy QCD.
It is necessary to develop a new {\it systematic-technology} to study the strongly coupled QCD. 
\\

\noindent
The standard model produces the Higgs boson \cite{Sirunyan:2017tqd} by the excitation of the Higgs field, which is called the Higgs mechanism. 
The Higgs mechanism is to study the perturbation in a low-energy state (or true vacuum-state) \cite{Coleman:1980aw, Stone:1975bd, Frampton:1976kf, Frampton:1976pb, Simon:1973yz}. 
This mechanism gives a natural way to interpret how to create a particle's mass.
Therefore, the observation in the Large Hadron Collider (LHC) shows that a perturbation method is a useful tool for studying the fundamental physics of our nature. 
However, the Higgs field has a non-zero value after the mechanism, and the non-zero value is inversely proportional to the square root of a coupling constant.
Although redefining the Higgs field for obtaining the value is not problematic, doing the {\it perturbation} becomes problematic. 
Because the unperturbed state is the eigenstate of the vanishing coupling constant case in the perturbation method, the unperturbed state is {\it not} a Fock state, labeled by the particle numbers \cite{Bargmann:1962zz}. 
The current method is to use an extremely weak coupling constant (but ignores the interacting terms) to work the perturbation.
In a perturbation study, the unperturbed state should be a Fock state.  
Hence the new skill is also necessary for the {\it weakly} coupled QFT.  
\\

\noindent
To solve the above issues, we study the {\it adaptive perturbation method} \cite{Weinstein:2005kw, Weinstein:2005kx}.
The adaptive perturbation decomposes a Hamiltonian by the diagonal elements of a Fock space and the non-diagonal parts of a Fock space \cite{Weinstein:2005kw, Weinstein:2005kx}. 
The perturbed term is the non-diagonal sector. 
Therefore, the perturbation parameter is {\it not} coupling constant \cite{Ma:2019pxd}. 
Because the unperturbed part is controlled by the diagonal elements of a Fock space as in the harmonic oscillator (but the adaptive perturbation method includes all diagonal elements), the unperturbed state is still a {\it Fock state}. 
To choose an unperturbed state with a {\it low} energy, the adaptive perturbation method introduces an {\it adaptive parameter} $\gamma$. 
It is convenient to give such the state from the variation of the parameter \cite{Weinstein:2005kw, Weinstein:2005kx}. 
The leading-order correction of the spectrum from the second-order perturbation provides a practical study, and it successfully approaches the numerical solution \cite{Curcio:2018, Ma:2020ipi}.
Hence the adaptive perturbation method seems to avoid the issue of the {\it weak-coupling perturbation}. 
The central question that we would like to address in this letter is the following: {\it Whether the adaptive perturbation method can apply to the Higgs field?}   
\\

\noindent
In this letter, we study the potential $\omega^2x^2/2+\lambda_1 x^4/6+\lambda_2 x^6/120$ because each perturbed term has an exact solution. 
The $\omega$ is frequency, $\lambda_1$ and $\lambda_2$ are coupling constants, and $x$ is the position operator.  
We show the analytical solution of the eigenenergy and $\langle x^2\rangle$ up to the next leading-order correction. 
The analytical formula shows a comparison to the numerical solution with a small deviation.
In particular, the inverted mass case ($\omega^2=-1$) provides direct evidence of the possible application of the Higgs field.

\section{Analytical Solution}
\label{sec:2}
\noindent
We introduce the adaptive perturbation method \cite{Weinstein:2005kw, Weinstein:2005kx} and show the analytical formula for the spectrum and the $\langle x^2\rangle$ up to the next leading-order correction. 
In the end, we demonstrate the accuracy of the analytical solution by comparing the perturbed solution to the numerical solution.

\subsection{Adaptive Perturbation Method}
\noindent
The main idea of the adaptive perturbation is to decide the decomposition of the Hamiltonian by whether the elements are in the diagonal places of the Fock state \cite{Weinstein:2005kw, Weinstein:2005kx}. 
To choose a suitable unperturbed state (with a low-energy), one introduces the adaptive parameter $\gamma$, which is allowed without changing the commutation relation $\lbrack p, x\rbrack=-i$ \cite{Weinstein:2005kw, Weinstein:2005kx}. 
The $p$ is the momentum operator. 
The $\gamma$ is introduced as \cite{Weinstein:2005kw, Weinstein:2005kx}: $x=(A_{\gamma}^{\dagger}+A_{\gamma})/\sqrt{2\gamma}$ and $p=i\sqrt{\gamma/2}(A_{\gamma}^{\dagger}-A_{\gamma})$, where $A_{\gamma}^{\dagger}$ is the creation operator, and $A_{\gamma}$ is the annihilation operator. 
The operators have the same relation as in the harmonic oscillator case, except for the dependence of the choice of $\gamma$, like the commutation relation $\lbrack A_{\gamma}, A_{\gamma}^{\dagger}\rbrack=1$ \cite{Weinstein:2005kw, Weinstein:2005kx}. 
Since the operators depend on the adaptive parameter, the vacuum state also depends ($A_{\gamma}|0_{\gamma}\rangle=0$) \cite{Weinstein:2005kw, Weinstein:2005kx}.
The adaptive parameter is just a scaling factor of the position operator.
Then the unperturbed part $H_0(\gamma)$ is replaced by the diagonal elements of the Fock space. 
The perturbed part $V(\gamma)$ is replaced by the non-diagonal elements of the Fock space. 
Here we study the Hamiltonian: 
\bea
H=H_0+V=\frac{p^2}{2}+\frac{\omega^2}{2}x^2+\frac{\lambda_1 }{6}x^4+\frac{\lambda_2}{120}x^6
\eea
because each perturbed term has an exact solution \cite{Ma:2019pxd}.

\subsection{Eigenenergy}
\noindent
Applying the time-independent perturbation to the adaptive perturbation method gives the same formula
\bea
&&E_n
\nn\\
&=&E_n^{(0)}+\sum_{k\neq n}\frac{|\langle k^{(0)}|V|n^{(0)}\rangle|^2}{E_n^{(0)}-E_{k, n}^{(0)}}
\nn\\
&&
+\sum_{k\neq n}\sum_{m\neq n}
\frac{\langle n^{(0)}|V|m^{(0)}\rangle\langle m^{(0)}|V|k^{(0)}\rangle\langle k^{(0)}|V|n^{(0)}\rangle}
{\big(E_n^{(0)}-E_{m, n}^{(0)}\big)\big(E_n^{(0)}-E_{k, n}^{(0)}\big)}+\cdots,
\eea
where $E_n^{(0)}$ is the $n$-th unperturbed eigenenergy, $|n^{(0)}\rangle$ is the $n$-th unperturbed eigenstate, and $E_{k, n}$ is the $k$-th unperturbed eigenenergy, calculated by the $n$-th unperturbed eigenstate's $\gamma$. 
The first-order term $\langle n^{(0)}|V|n^{(0)}\rangle$ vanishes due to that $V$ is a non-diagonal element of the Fock space. 
The adaptive parameter $\gamma$ is determined by minimizing the unperturbed spectrum for the parameter \cite{Ma:2019pxd, Curcio:2018}   
\bea
&&E_n^{(0)}
\nn\\
&=&\frac{\gamma}{4}(2n+1)+\frac{\omega^2}{4\gamma}(2n+1)
\nn\\
&&+\frac{\lambda_1}{4\gamma^2}\bigg(n^2+n+\frac{1}{2}\bigg)
+\frac{\lambda_2}{4\gamma^3}\bigg(\frac{1}{12}n^3+\frac{29}{240}n^2+\frac{1}{6}n+\frac{1}{16}\bigg),
\nn\\
\eea 
in which the $\gamma$ is positive, and it satisfies the algebra equation \cite{Ma:2019pxd}
\bea
\gamma^4-\omega^2\gamma^2-\lambda_1\frac{2n^2+2n+1}{2n+1}\gamma
-\frac{\lambda_2}{80}\frac{20n^3+29n^2+40n+15}{2n+1}
=0.
\nn\\
\eea
Here we use the value of the $\gamma$ in the higher-order calculation as in the solvable part. 
Because the adaptive parameter depends on the value of $n$, it is hard to guarantee whether the adaptive perturbation method is practical (although the choice of the adaptive parameter should not affect the result if we do the perturbation to all-orders). 
However, the second-order perturbation is already enough to show a small deviation to the numerical solution \cite{Curcio:2018, Ma:2020ipi}.  
\\

\subsubsection{2nd-Order and 3rd-Order}
\noindent
To write the perturbed spectrum conveniently, we introduce the below parameters:
\bea
T_1&\equiv&\frac{\lambda_2}{960\gamma^3};
\nn\\
T_2&\equiv&\frac{\lambda_1}{24\gamma^2}+\frac{\lambda_2}{320\gamma^3}(2n+5);
\nn\\
T_3&\equiv&-\frac{\gamma}{4}
+\frac{\omega^2}{4\gamma}
+\frac{\lambda_1}{12\gamma^2}(2n+3)
+\frac{\lambda_2}{64\gamma^3}(n^2+3n+3);
\nn\\
T_4&\equiv&-\frac{\gamma}{4}
+\frac{\omega^2}{4\gamma}
+\frac{\lambda_1}{12\gamma^2}(2n-1)
+\frac{\lambda_2}{64\gamma^3}(n^2-n+1);
\nn\\
T_5&\equiv&\frac{\lambda_1}{24\gamma^2}+\frac{\lambda_2}{320\gamma^3}(2n-3);
\nn\\
T_6&\equiv&T_1;
\nn\\
T_7&\equiv&-\frac{\gamma}{4}
+\frac{\omega^2}{4\gamma}
+\frac{\lambda_1}{12\gamma^2}(2n+11)
+\frac{\lambda_2}{64\gamma^3}(n^2+11n+31);
\nn\\
T_8&\equiv&\frac{\lambda_1}{24\gamma^2}
+\frac{\lambda_2}{320\gamma^3}(2n+9);
\nn\\
T_9&\equiv&-\frac{\gamma}{4}
+\frac{\omega^2}{4\gamma}
+\frac{\lambda_1}{12\gamma^2}(2n+7)
+\frac{\lambda_2}{64\gamma^3}(n^2+7n+13);
\nn\\
T_{10}&\equiv&\frac{\lambda_1}{24\gamma^2}
+\frac{\lambda_2}{320\gamma^3}(2n+1);
\nn\\
T_{11}&\equiv&-\frac{\gamma}{4}
+\frac{\omega^2}{4\gamma}
+\frac{\lambda_1}{12\gamma^2}(2n-5)
+\frac{\lambda_2}{64\gamma^3}(n^2-5n+7);
\nn\\
T_{12}&\equiv&\frac{\lambda_1}{24\gamma^2}+\frac{\lambda_2}{320\gamma^3}(2n-7);
\nn\\
T_{13}&\equiv&-\frac{\gamma}{4}
+\frac{\omega^2}{4\gamma}
+\frac{\lambda_1}{12\gamma^2}(2n-9)
+\frac{\lambda_2}{64\gamma^3}(n^2-9n+21).
\eea
The necessary of the transition energy is given as in the following \cite{Ma:2019pxd}:
\bea
&&E_n^{(0)}(\gamma)-E_{n+6, n}^{(0)}(\gamma)
\nn\\
&=&-3\gamma
-\frac{3\omega^2}{\gamma}
-\frac{3\lambda_1}{2\gamma^2}(2n+7)
+\frac{\lambda_2}{4\gamma^3}\bigg(-\frac{3}{2}(n^2+6n+12)-\frac{29}{20}(n+3)-1\bigg);
\nn\\
&&E_n^{(0)}(\gamma)-E_{n+4, n}^{(0)}(\gamma)
\nn\\
&=&-2\gamma
-\frac{2\omega^2}{\gamma}
-\frac{\lambda_1}{\gamma^2}(2n+5)
-\frac{\lambda_2}{4\gamma^3}\bigg\lbrack\bigg(n^2+4n+\frac{16}{3}\bigg)+\frac{29}{30}(n+2)+\frac{2}{3}\bigg\rbrack;
\nn\\
&&E_n^{(0)}(\gamma)-E_{n+2, n}^{(0)}(\gamma)
\nn\\
&=&-\gamma
-\frac{\omega^2}{\gamma}
-\frac{\lambda_1}{2\gamma^2}(2n+3)
-\frac{\lambda_2}{4\gamma^3}\bigg\lbrack\bigg(\frac{1}{2}n^2+n+\frac{2}{3}\bigg)+\frac{29}{60}(n+1)+\frac{1}{3}\bigg\rbrack;
\nn\\
&&E_n^{(0)}(\gamma)-E_{n-2, n}^{(0)}(\gamma)
\nn\\
&=&\gamma
+\frac{\omega^2}{\gamma}
+\frac{\lambda_1}{2\gamma^2}(2n-1)
+\frac{\lambda_2}{4\gamma^3}\bigg\lbrack\bigg(\frac{1}{2}n^2-n+\frac{2}{3}\bigg)+\frac{29}{60}(n-1)+\frac{1}{3}\bigg\rbrack;
\nn\\
&&E_n^{(0)}(\gamma)-E_{n-4, n}^{(0)}(\gamma)
\nn\\
&=&2\gamma
+\frac{2\omega^2}{\gamma}
+\frac{\lambda_1}{\gamma^2}(2n-3)
+\frac{\lambda_2}{4\gamma^3}\bigg\lbrack\bigg(n^2-4n+\frac{16}{3}\bigg)+\frac{29}{30}(n-2)+\frac{2}{3}\bigg\rbrack;
\nn\\
&&E_n^{(0)}(\gamma)-E_{n-6, n}^{(0)}(\gamma)
\nn\\
&=&3\gamma
+\frac{3\omega^2}{\gamma}
+\frac{3\lambda_1}{2\gamma^2}(2n-5)
+\frac{\lambda_2}{4\gamma^3}\bigg(\frac{3}{2}(n^2-6n+12)+\frac{29}{20}(n-3)+1\bigg).
\nn\\
\eea

\noindent
The second-order perturbation gives \cite{Ma:2020ipi}:
\bea
&&E_n(\gamma)_2
\nn\\
&=&
E_n^{(0)}(\gamma)+\sum_{k\neq n}\frac{|\langle k^{(0)}|V(\gamma)|n^{(0)}\rangle|^2}{E_n^{(0)}(\gamma)-E_{k, n}^{(0)}(\gamma)}
\nn\\
&=&E_n^{(0)}
\nn\\
&&+\frac{T_1^2}{E_n^{(0)}-E_{n+6, n}^{(0)}}(n+1)(n+2)(n+3)(n+4)(n+5)(n+6)
\nn\\
&&
+\frac{T_2^2}{E_n^{(0)}-E_{n+4, n}^{(0)}}(n+1)(n+2)(n+3)(n+4)
\nn\\
&&
+\frac{T_3^2}{E_n^{(0)}-E_{n+2, n}^{(0)}}(n+1)(n+2)
\nn\\
&&
+\frac{T_4^2}{E_n^{(0)}-E_{n-2, n}^{(0)}}(n-1)n
\nn\\
&&
+\frac{T_5^2}{E_n^{(0)}-E_{n-4, n}^{(0)}}(n-3)(n-2)(n-1)n
\nn\\
&&
+\frac{T_6^2}{E_n^{(0)}-E_{n-6, n}^{(0)}}(n-5)(n-4)(n-3)(n-2)(n-1)n.
\eea
The third-order perturbation gives:
\bea
&&E_n(\gamma)_3
\nn\\
&=&
E_n(\gamma)_2
+\sum_{k\neq n}\sum_{m\neq n}
\frac{\langle n^{(0)}|V(\gamma)|m^{(0)}\rangle\langle m^{(0)}|V(\gamma)|k^{(0)}\rangle\langle k^{(0)}|V(\gamma)|n^{(0)}\rangle}
{\big(E_n^{(0)}(\gamma)-E_{m, n}^{(0)}(\gamma)\big)\big(E_n^{(0)}(\gamma)-E_{k, n}^{(0)}(\gamma)\big)}
\nn\\
&=&E_n(\gamma)_2
\nn\\
&&+(n+1)(n+2)(n+3)(n+4)(n+5)(n+6)
\nn\\
&&\times\bigg(\frac{T_1T_2T_7}{\big(E_n^{(0)}-E_{n+6, n}^{(0)}\big)\big(E_n^{(0)}-E_{n+4, n}^{(0)}\big)}
+\frac{T_1T_3T_8}{\big(E_n^{(0)}-E_{n+6, n}^{(0)}\big)\big(E_n^{(0)}-E_{n+2, n}^{(0)}\big)}\bigg)
\nn\\
&&+(n+1)(n+2)(n+3)(n+4)
\nn\\
&&\times\bigg(
\frac{(n+5)(n+6)T_1T_2T_7}{\big(E_n^{(0)}-E_{n+4, n}^{(0)}\big)\big(E_n^{(0)}-E_{n+6, n}^{(0)}\big)}
+\frac{T_2T_3T_9}{\big(E_n^{(0)}-E_{n+4, n}^{(0)}\big)\big(E_n^{(0)}-E_{n+2, n}^{(0)}\big)}
\nn\\
&&+\frac{(n-1)nT_2T_4T_6}{\big(E_n^{(0)}-E_{n+4, n}^{(0)}\big)\big(E_n^{(0)}-E_{n-2, n}^{(0)}\big)}\bigg)
\nn\\
&&+(n+1)(n+2)
\nn\\
&&\times\bigg(
\frac{(n+3)(n+4)(n+5)(n+6)T_1T_3T_8}{\big(E_n^{(0)}-E_{n+2, n}^{(0)}\big)\big(E_n^{(0)}-E_{n+6, n}^{(0)}\big)}
+
\frac{(n+3)(n+4)T_2T_3T_9}{\big(E_n^{(0)}-E_{n+2, n}^{(0)}\big)\big(E_n^{(0)}-E_{n+4, n}^{(0)}\big)}
\nn\\
&&
+
\frac{(n-1)nT_3T_4T_{10}}{\big(E_n^{(0)}-E_{n+2, n}^{(0)}\big)\big(E_n^{(0)}-E_{n-2, n}^{(0)}\big)}
+
\frac{(n-3)(n-2)(n-1)nT_3T_5T_6}{\big(E_n^{(0)}-E_{n+2, n}^{(0)}\big)\big(E_n^{(0)}-E_{n-4, n}^{(0)}\big)}\bigg)
\nn\\
&&+(n-1)n
\nn\\
&&\times
\bigg(\frac{(n+1)(n+2)(n+3)(n+4)T_1T_2T_4}{\big(E_n^{(0)}-E_{n-2, n}^{(0)}\big)\big(E_n^{(0)}-E_{n+4, n}^{(0)}\big)}
+\frac{(n+1)(n+2)T_3T_4T_{10}}{\big(E_n^{(0)}-E_{n-2, n}^{(0)}\big)\big(E_n^{(0)}-E_{n+2, n}^{(0)}\big)}
\nn\\
&&
+\frac{(n-3)(n-2)T_4T_5T_{11}}{\big(E_n^{(0)}-E_{n-2, n}^{(0)}\big)\big(E_n^{(0)}-E_{n-4, n}^{(0)}\big)}
+\frac{(n-5)(n-4)(n-3)(n-2)T_4T_6T_{12}}{\big(E_n^{(0)}-E_{n-2, n}^{(0)}\big)\big(E_n^{(0)}-E_{n-6, n}^{(0)}\big)}\bigg)
\nn\\
&&+(n-3)(n-2)(n-1)n
\nn\\
&&\times\bigg(\frac{(n+1)(n+2)T_1T_3T_5}{\big(E_n^{(0)}-E_{n-4, n}^{(0)}\big)\big(E_n^{(0)}-E_{n+2, n}^{(0)}\big)}
+\frac{T_4T_5T_{11}}{\big(E_n^{(0)}-E_{n-4, n}^{(0)}\big)\big(E_n^{(0)}-E_{n-2, n}^{(0)}\big)}
\nn\\
&&
+\frac{(n-5)(n-4)T_5T_6T_{13}}{\big(E_n^{(0)}-E_{n-4, n}^{(0)}\big)\big(E_n^{(0)}-E_{n-6, n}^{(0)}\big)}\bigg)
\nn\\
&&+(n-5)(n-4)(n-3)(n-2)(n-1)n
\nn\\
&&\times\bigg(
\frac{T_4T_6T_{12}}{\big(E_n^{(0)}-E_{n-6, n}^{(0)}\big)\big(E_n^{(0)}-E_{n-2, n}^{(0)}\big)}
+\frac{T_5T_6T_{13}}{\big(E_n^{(0)}-E_{n-6, n}^{(0)}\big)\big(E_n^{(0)}-E_{n-4, n}^{(0)}\big)}\bigg).
\nn\\
\eea

\subsubsection{Numerical Solution}
We first demonstrate the accuracy of the analytical formula for the $\omega^2=1$ and $(\lambda_1, \lambda_2)=(16, 0); (16, 256)$ in Tables \ref{c1160} and \ref{c116256}. 
\begin{table}[!htb]
\centering
\begin{tabular}{ |m{1em} | m{2cm}| m{2cm}|m{4cm}|m{3cm}|m{3cm}|} 
\hline
\textbf{$n$} & \textbf{$E_n(\gamma)_{2}$} & \textbf{$E_n(\gamma)_{3}$}&\textbf{Numerical Solution}&\textbf{Deviation} 1&\textbf{Deviation} 2 \\ 
\hline
$0$ & 1.0292&1.0292 &1.0268&0.2337\%& 0.2337\%\\ 
\hline
$1$ & 3.5762&3.5762 &3.5721&0.1147\%& 0.1147\%\\ 
\hline
$2$ & 6.8789&6.8698 &6.865&0.2024\%& 0.0699\%\\ 
\hline
$3$ & 10.6461&10.6216 &10.6141&0.3014\%& 0.0706\%\\ 
\hline
$4$ & 14.7802&14.7385 &14.7287&0.3496\%& 0.0665\%\\ 
\hline
$5$ & 19.224&19.1638 &19.1514&0.379\%& 0.0647\%\\ 
\hline
$6$ & 23.9384&23.8587 &23.8437&0.3971\%& 0.0629\%\\ 
\hline
$7$ & 28.8951&28.7949 &28.777&0.4103\%& 0.0622\%\\ 
\hline
\end{tabular}
\caption{The comparison between the perturbation and numerical solutions for the $\omega^2=1$, $\lambda_1=16$, and $\lambda_2=0$.}
\label{c1160}
\end{table}
\begin{table}[!htb]
\centering
\begin{tabular}{ |m{1em} | m{2cm}| m{2cm}|m{4cm}|m{3cm}|m{3cm}|} 
\hline
\textbf{$n$} & \textbf{$E_n(\gamma)_{2}$} & \textbf{$E_n(\gamma)_{3}$}&\textbf{Numerical Solution}&\textbf{Deviation} 1&\textbf{Deviation} 2 \\ 
\hline
$0$ & 1.1681&1.172 &1.1599&0.7069\%& 1.0431\%\\ 
\hline
$1$ & 4.1655&4.1761 &4.1545&0.2647\%& 0.5199\%\\ 
\hline
$2$ & 8.2973&8.2828 &8.2622&0.4248\%& 0.2493\%\\ 
\hline
$3$ & 13.2538&13.19 &13.1621&0.6966\%& 0.2119\%\\ 
\hline
$4$ & 18.8898&18.7598 &18.7216&0.8984\%& 0.204\%\\ 
\hline
$5$ & 25.12&24.9111 &24.8604&1.0442\%& 0.2039\%\\ 
\hline
$6$ & 31.8858&31.5873 &31.522&1.1541\%& 0.2071\%\\ 
\hline
$7$ & 39.1433&38.7459 &38.6639&1.2399\%& 0.212\%\\ 
\hline
\end{tabular}
\caption{The comparison between the perturbation and numerical solutions for the $\omega^2=1$, $\lambda_1=16$, and $\lambda_2=256$.}
\label{c116256}
\end{table}

\noindent
The accuracy in the strongly coupled region is lower than or around 1\% within the third-order perturbation. 
Therefore, the perturbation study gives an accurate analytical-formula to the strong coupling region, and indeed, other coupling regions also show so. 
The $\textbf{Deviation}\ 1$ is defined as the deviation of the leading-order correction from the numerical solution. 
The $\textbf{Deviation}\ 2$ is defined as the deviation of the next leading-order correction from the numerical solution in all Tables. 
\\

\noindent
The Hamiltonian in the numerical study is defined by the discretized kinetic energy $(p^2/2)\psi\rightarrow-(\psi_{j+1}-2\psi_j+\psi_{j-1})/(2a^2)$,
where $\psi_j$ is the eigenfunction at the site $x_j$ in the discrete theory, and $a$ is the lattice spacing. The lattice index is labeled by $j=1, 2, \cdots, n$, where $n$ is the number of lattice points. 
The discrete system has $n+1$ lattice points with a lattice size $2L$ and the periodic boundary condition as the below:
\bea
-L\le x_{j}\le L; \qquad x_{0}=-L; \qquad x_{j+1}\equiv x_{j}+a; \qquad
 \psi_{0}\equiv \psi_{n}; \qquad  2L=na.
 \nn\\
\eea 
The numerical parameters in this letter are chosen as that: $L=8$ and $n=16384$.
\\

\noindent
In the end, we show the accurate study from $\omega^2=0$ in Tables \ref{c0160} and \ref{c016256}. 
When coupling constants vanish, the unperturbed state is not a Fock state. 
Therefore, the region cannot be applied to the weak-coupling perturbation. 
Therefore, the accurate result should demonstrate the applicability of adaptive perturbation method to the similar problem of Higgs field before we give a more direct evidence from the inverted mass term. 
\begin{table}[!htb]
\centering
\begin{tabular}{ |m{1em} | m{2cm}| m{2cm}|m{4cm}|m{3cm}|m{3cm}|} 
\hline
\textbf{$n$} & \textbf{$E_n(\gamma)_{2}$} & \textbf{$E_n(\gamma)_{3}$}&\textbf{Numerical Solution}&\textbf{Deviation} 1&\textbf{Deviation} 2 \\ 
\hline
$0$ & 0.9299&0.9299 & 0.9263&0.3886\%& 0.3886\%\\ 
\hline
$1$ & 3.3249&3.3249 & 3.3193&0.1687\%& 0.1687\%\\ 
\hline
$2$ & 6.5305&6.5194 & 6.5131&0.2671\%& 0.0967\%\\ 
\hline
$3$ & 10.2112&10.1822 & 10.1726&0.3794\%& 0.0943\%\\ 
\hline
$4$ & 14.2664&14.2181 & 14.206&0.4251\%& 0.0851\%\\ 
\hline
$5$ & 18.6366&18.5683 & 18.5534&0.4484\%& 0.0803\%\\ 
\hline
$6$ & 23.2818&23.1925 & 23.1747&0.4621\%& 0.0768\%\\ 
\hline
$7$ & 28.1727&28.0616 & 28.0406&0.4711\%& 0.0748\%\\ 
\hline
\end{tabular}
\caption{The comparison between the perturbation and numerical solutions for the $\omega^2=0$, $\lambda_1=16$, and $\lambda_2=0$.}
\label{c0160}
\end{table}
\begin{table}[!htb]
\centering
\begin{tabular}{ |m{1em} | m{2cm}| m{2cm}|m{4cm}|m{3cm}|m{3cm}|} 
\hline
\textbf{$n$} & \textbf{$E_n(\gamma)_{2}$} & \textbf{$E_n(\gamma)_{3}$}&\textbf{Numerical Solution}&\textbf{Deviation} 1&\textbf{Deviation} 2 \\ 
\hline
$0$ & 1.0864&1.0913 & 1.0757&0.9947\%& 1.4502\%\\ 
\hline
$1$ & 3.964&3.976 &3.9499&0.3569\%& 0.6607\%\\ 
\hline
$2$ & 8.0329&8.0165 &7.9916&0.5167\%& 0.3115\%\\ 
\hline
$3$ & 12.9395&12.8694 &12.8359&0.8071\%& 0.2609\%\\ 
\hline
$4$ & 18.5321&18.3916 &18.3469&1.0094\%& 0.2436\%\\ 
\hline
$5$ & 24.7232&24.5 &24.4421&1.15\%& 0.2368\%\\ 
\hline
$6$ & 31.4532&31.1371 &31.0639&1.2532\%& 0.2356\%\\ 
\hline
$7$ & 38.6774&38.2595 &38.1688&1.3325\%& 0.2376\%\\ 
\hline
\end{tabular}
\caption{The comparison between the perturbation and numerical solutions for the $\omega^2=0$, $\lambda_1=16$, and $\lambda_2=256$.}
\label{c016256}
\end{table}

\subsection{$\langle x^2\rangle$}
\noindent
The complete information of a quantum system contains the eigenenergy and also the eigenstates. 
Usually, it is harder to obtain an accurate eigenstate than the eigenvalue. 
When the value of the adaptive parameter is determined by the minimization of energy, it possibly only guarantees that the suitable perturbed state for calculating eigenenergy, but it may not be adaptive enough for the correlation functions. 
In QFT, we are interested in the correlation functions due to the motivation from experiments. 
Hence it is important to show the applicability of correlation functions in Quantum Mechanics (QM) before we apply the adaptive perturbation method to QFT.
To demonstrate the suitability of the eigenstate in the adaptive perturbation method,  we show the perturbation of $\langle x^2\rangle$ up to the second-order with a small deviation to the numerical solution for the $\omega^2=1, 0; (\lambda_1, \lambda_2)=(16, 0)$ in Tables \ref{c21160} and \ref{c20160}. 
\begin{table}[!htb]
\centering
\begin{tabular}{ |m{1em} | m{2cm}| m{2cm}|m{4cm}|m{3cm}|m{3cm}|} 
\hline
\textbf{$n$} & \textbf{$S_n(\gamma)_{1}$} & \textbf{$S_n(\gamma)_{2}$}&\textbf{Numerical Solution}&\textbf{Deviation} 1&\textbf{Deviation} 2 \\ 
\hline
$0$ & 0.1885& 0.1953&0.1951&3.3828\%& 0.1025\%\\ 
\hline
$1$ &0.484& 0.4968 &0.4954&2.3011\%& 0.2825\%\\ 
\hline
$2$ &0.6869 &0.6949 &0.6951&1.1796\%& 0.0287\%\\ 
\hline
$3$ &0.8629 &0.8703 &0.8742&1.2926\%& 0.4461\%\\ 
\hline
$4$ &1.0226 &1.0302 &1.0366&1.3505\%& 0.6174\%\\ 
\hline
$5$ &1.1708 &1.179 &1.1874&1.398\%& 0.7074\%\\ 
\hline
$6$ &1.3103 &1.3191 &1.3292&1.4219\%& 0.7598\%\\ 
\hline
$7$ &1.4428 &1.4523 &1.464&1.448\%& 0.7991\%\\ 
\hline
\end{tabular}
\caption{The comparison between the perturbation and numerical solutions for the $\omega^2=1$, $\lambda_1=16$, and $\lambda_2=0$.}
\label{c21160}
\end{table}
\begin{table}[!htb]
\centering
\begin{tabular}{ |m{1em} | m{2cm}| m{2cm}|m{4cm}|m{3cm}|m{3cm}|} 
\hline
\textbf{$n$} & \textbf{$S_n(\gamma)_{1}$} & \textbf{$S_n(\gamma)_{2}$}&\textbf{Numerical Solution}&\textbf{Deviation} 1&\textbf{Deviation} 2 \\ 
\hline
$0$ &0.1984 &0.2073 & 0.2072&4.2471\%& 0.0482\%\\ 
\hline
$1$ &0.502 &0.5176 & 0.516&2.7131\%& 0.31\%\\ 
\hline
$2$ &0.703 &0.7118 & 0.7124&1.3194\%& 0.0842\%\\ 
\hline
$3$ &0.8784 &0.8864 & 0.8916&1.4804\%& 0.5832\%\\ 
\hline
$4$ &1.0378 &1.0461 & 1.054&1.537\%& 0.7495\%\\ 
\hline
$5$ &1.1859 &1.1947 & 1.2048&1.5687\%& 0.8383\%\\ 
\hline
$6$ &1.3253 &1.3347 & 1.3466&1.5817\%& 0.8837\%\\ 
\hline
$7$ &1.4578 &1.4679 & 1.4814&1.593\%& 0.9113\%\\ 
\hline
\end{tabular}
\caption{The comparison between the perturbation and numerical solutions for the $\omega^2=0$, $\lambda_1=16$, and $\lambda_2=0$.}
\label{c20160}
\end{table}
\\

\noindent
The perturbed eigenstate is
\bea
&&
|n\rangle
\nn\\
&=&|n^{(0)}\rangle
+\sum_{k\neq n}\frac{\langle k^{(0)}|V|n^{(0)}\rangle}{E_n^{(0)}-E_{k, n}^{(0)}}|k^{(0)}\rangle
\nn\\
&&
+\bigg(\sum_{k\neq n}\sum_{l\neq n}\frac{\langle k^{(0)}|V|l^{(0)}\rangle\langle l^{(0)}|V|n^{(0)}\rangle}{\big(E_n^{(0)}-E_{k, n}^{(0)}\big)\big(E_n^{(0)}-E_{l, n}^{(0)}\big)}|k^{(0)}\rangle
\nn\\
&&
-\frac{1}{2}\sum_{k\neq n}\frac{\langle n^{(0)}|V|k^{(0)}\rangle\langle k^{(0)}|V|n^{(0)}\rangle}{\big(E_n^{(0)}-E_{k, n}^{(0)}\big)^2}|n^{(0)}\rangle\bigg)
+\cdots.
\eea
The leading-order is $|n^{(0)}\rangle$, the leading-order correction is 
\bea
|n^{(1)}\rangle=\sum_{k\neq n}\frac{\langle k^{(0)}|V|n^{(0)}\rangle}{E_n^{(0)}-E_{k, n}^{(0)}}|k^{(0)}\rangle,
\eea
and the next leading-order correction is
\bea
&&|n^{(2)}\rangle
\nn\\
&=&\sum_{k\neq n}\sum_{l\neq n}\frac{\langle k^{(0)}|V|l^{(0)}\rangle\langle l^{(0)}|V|n^{(0)}\rangle}{\big(E_n^{(0)}-E_{k, n}^{(0)}\big)\big(E_n^{(0)}-E_{l, n}^{(0)}\big)}|k^{(0)}\rangle
\nn\\
&&
-\frac{1}{2}\sum_{k\neq n}\frac{\langle n^{(0)}|V|k^{(0)}\rangle\langle k^{(0)}|V|n^{(0)}\rangle}{\big(E_n^{(0)}-E_{k, n}^{(0)}\big)^2}|n^{(0)}\rangle.
\eea
\\

\noindent
The $\langle x^2\rangle$ up to the first-order correction is given by:
\bea
&&
S_n(\gamma)_1
\nn\\
&=&
\langle n^{(0)}|x^2|n^{(0)}\rangle
+2\langle n^{(1)}|x^2|n^{(0)}\rangle
\nn\\
&=&\frac{1}{2\gamma}(2n+1)
+\frac{1}{\gamma}\bigg((n+1)(n+2)\frac{T_3}{E_n^{(0)}-E_{n+2, n}^{(0)}}+(n-1)n\frac{T_4}{E_n^{(0)}-E_{n-2, n}^{(0)}}\bigg).
\nn\\
\eea
\\

\noindent
The $\langle x^2\rangle$ up to the second-order correction is given by:
\bea
&&S_n(\gamma)_2
\nn\\
&=&S_n(\gamma)_1+2\langle n^{(2)}|x^2|n^{(0)}\rangle
+\langle n^{(1)}|x^2|n^{(1)}\rangle
\nn\\
&=&
S_n(\gamma)_1
\nn\\
&&+
\frac{1}{\gamma}\bigg\lbrack (n+1)(n+2)\bigg(\frac{(n-3)(n-2)(n-1)nT_5T_6}{\big(E_n^{(0)}-E_{n+2, n}^{(0)}\big)\big(E_n^{(0)}-E_{n-4, n}^{(0)}\big)}
\nn\\
&&
+\frac{(n-1)nT_4T_{10}}{\big(E_n^{(0)}-E_{n+2, n}^{(0)}\big)\big(E_n^{(0)}-E_{n-2, n}^{(0)}\big)}
+\frac{(n+3)(n+4)T_2T_9}{\big(E_n^{(0)}-E_{n+2, n}^{(0)}\big)\big(E_n^{(0)}-E_{n+4, n}^{(0)}\big)}
\nn\\
&&
+\frac{(n+3)(n+4)(n+5)(n+6)T_1T_8}{\big(E_n^{(0)}-E_{n+2, n}^{(0)}\big)\big(E_n^{(0)}-E_{n+6, n}^{(0)}\big)}\bigg)
\nn\\
&&
+(n-1)n\bigg(\frac{(n-5)(n-4)(n-3)(n-2)T_6T_{12}}{\big(E_n^{(0)}-E_{n-2, n}^{(0)}\big)\big(E_n^{(0)}-E_{n-6, n}^{(0)}\big)}
+\frac{(n-3)(n-2)T_5T_{11}}{\big(E_n^{(0)}-E_{n-2, n}^{(0)}\big)\big(E_n^{(0)}-E_{n-4, n}^{(0)}\big)}
\nn\\
&&+\frac{(n+1)(n+2)T_3T_{10}}{\big(E_n^{(0)}-E_{n-2, n}^{(0)}\big)\big(E_n^{(0)}-E_{n+2, n}^{(0)}\big)}
+\frac{(n+1)(n+2)(n+3)(n+4)T_1T_2}{\big(E_n^{(0)}-E_{n-2, n}^{(0)}\big)\big(E_n^{(0)}-E_{n-4, n}^{(0)}\big)}\bigg)
\nn\\
&&
+\frac{(n+1)(n+2)(n+3)(n+4)T_2T_3}{\big(E_n^{(0)}-E_{n+4, n}^{(0)}\big)\big(E_n^{(0)}-E_{n+2, n}^{(0)}\big)}
\nn\\
&&
+\frac{(n+1)(n+2)(n+3)(n+4)(n+5)(n+6)T_1T_2}{\big(E_n^{(0)}-E_{n+6, n}^{(0)}\big)\big(E_n^{(0)}-E_{n+4, n}^{(0)}\big)}
+\frac{(n-3)(n-2)(n-1)nT_4T_5}{\big(E_n^{(0)}-E_{n-2, n}^{(0)}\big)\big(E_n^{(0)}-E_{n-4, n}^{(0)}\big)}
\nn\\
&&
+\frac{(n-5)(n-4)(n-3)(n-2)(n-1)nT_5T_6}{\big(E_n^{(0)}-E_{n-4, n}^{(0)}\big)\big(E_n^{(0)}-E_{n-6, n}^{(0)}\big)}
\nn\\
&&
-\frac{6(n-5)(n-4)(n-3)(n-2)(n-1)nT_6^2}{\big(E_n^{(0)}-E_{n-6, n}^{(0)}\big)^2}
-\frac{4(n-3)(n-2)(n-1)nT_5^2}{\big(E_n^{(0)}-E_{n-4, n}^{(0)}\big)^2}
\nn\\
&&
-\frac{2(n-1)nT_4^2}{\big(E_n^{(0)}-E_{n-2, n}^{(0)}\big)^2}
+\frac{2(n+1)(n+2)T_3^2}{\big(E_n^{(0)}-E_{n+2, n}^{(0)}\big)^2}
+\frac{4(n+1)(n+2)(n+3)(n+4)T_2^2}{\big(E_n^{(0)}-E_{n+4, n}^{(0)}\big)^2}
\nn\\
&&
+\frac{6(n+1)(n+2)(n+3)(n+4)(n+5)(n+6)T_1^2}{\big(E_n^{(0)}-E_{n+6, n}^{(0)}\big)^2}
\bigg\rbrack.
\nn\\
\eea

\section{Inverted Mass Term}
\label{sec:3}
\noindent
Now we discuss the analytical formula for the inverted mass term ($\omega^2=-1$). 
For an oscillator, the eigenenergy is proportional to $\omega$. 
Therefore, the analytical continuation of the $\omega$ from the real-valued number to the imaginary-valued number should be failed for the weak-coupling perturbation. 
The Hamiltonian is bounded from below for any positive value of the coupling constants. 
Therefore, the analytical continuation should not be problematic. 
Indeed, the real problem is the perturbation method.
The perturbation relies on the existence of a discrete spectrum or a Fock state at the leading-order. 
The adaptive perturbation method \cite{Weinstein:2005kw, Weinstein:2005kx} includes the coupling constants in the leading-order perturbation \cite{Weinstein:2005kw, Weinstein:2005kx}. 
Hence the inverted mass case still has a discrete spectrum at the unperturbed level. 
For an inverted oscillator case, the eigenenergy can be smaller than zero.
If the adaptive perturbation method is compatible with the analytical continuation, we can determine what is the critical value of the coupling constants for having a positive definite spectrum in the model. 
Here we turn off the $\lambda_2$ for convenience. 
The third-order calculation shows the critical value lies on $0.8344-0.8345$ with a deviation of less than 10\% from the true critical-value $0.9072-0.9073$ (numerical value). 
When one chooses the $\lambda_1=0$, the system loses a discrete spectrum. 
Hence the 10\% deviation is not strange in the weak-coupling region, and one can find that the perturbation result approaches the numerical value by introducing the higher-order terms. 
\\

\noindent
As we discussed above, the study of the inverted mass case is non-trivial because the weak-coupling perturbation is failed in a strongly coupled region as in the usual situation and also in a weakly coupled region. 
To give a concrete evidence for the application of the adaptive perturbation to the inverted mass term, we show the eigenenergy and $\langle x^2\rangle$ for the case $(\lambda_1, \lambda_2)=(16, 0)$ in Tables \ref{c-1160} and \ref{c2-1160} respectively. 
\begin{table}[!htb]
\centering
\begin{tabular}{ |m{1em} | m{2cm}| m{2cm}|m{4cm}|m{3cm}|m{3cm}|} 
\hline
\textbf{$n$} & \textbf{$E_n(\gamma)_{2}$} & \textbf{$E_n(\gamma)_{3}$}&\textbf{Numerical Solution}&\textbf{Deviation} 1&\textbf{Deviation} 2 \\ 
\hline
$0$ & 0.8247&0.8247 &0.8193&0.659\%& 0.659\%\\ 
\hline
$1$ & 3.0633&3.0633 &3.05574&0.2474\%& 0.2474\%\\ 
\hline
$2$ & 6.1743&6.1608 &6.1525&0.3543\%& 0.1349\%\\ 
\hline
$3$ & 9.7691&9.7346 &9.7223&0.4813\%& 0.1265\%\\ 
\hline
$4$ & 13.7455&13.6896 &13.6745&0.5192\%& 0.1104\%\\ 
\hline
$5$ & 18.0423&17.9646 &17.9466&0.5332\%& 0.1002\%\\ 
\hline
$6$ & 22.6183&22.5181 &22.497&0.5391\%& 0.0937\%\\ 
\hline
$7$ & 27.4434&27.32 &27.2955&0.5418\%& 0.0897\%\\ 
\hline
\end{tabular}
\caption{The comparison between the perturbation and numerical solutions for the $\omega^2=-1$, $\lambda_1=16$, and $\lambda_2=0$.}
\label{c-1160}
\end{table}  
\begin{table}[!htb]
\centering
\begin{tabular}{ |m{1em} | m{2cm}| m{2cm}|m{4cm}|m{3cm}|m{3cm}|} 
\hline
\textbf{$n$} & \textbf{$S_n(\gamma)_{1}$} & \textbf{$S_n(\gamma)_{2}$}&\textbf{Numerical Solution}&\textbf{Deviation} 1&\textbf{Deviation} 2 \\ 
\hline
$0$ &0.2094 &0.221 & 0.2211&5.2917\%& 0.0452\%\\ 
\hline
$1$ &0.5215 &0.5405 & 0.5386&3.1748\%& 0.3527\%\\ 
\hline
$2$ &0.7193 &0.7288 & 0.73&1.4657\%& 0.1643\%\\ 
\hline
$3$ &0.8939 &0.9025 & 0.9095&1.7152\%& 0.7696\%\\ 
\hline
$4$ &1.053 &1.0619 & 1.0718&1.754\%& 0.9236\%\\ 
\hline
$5$ &1.2009 &1.2104 & 1.2225&1.7668\%& 0.9897\%\\ 
\hline
$6$ &1.3403 &1.3503 & 1.3643&1.7591\%& 1.0261\%\\ 
\hline
$7$ &1.4727 &1.4834 & 1.4991&1.761\%& 1.0472\%\\ 
\hline
\end{tabular}
\caption{The comparison between the perturbation and numerical solutions for the $\omega^2=-1$, $\lambda_1=16$, and $\lambda_2=0$.}
\label{c2-1160}
\end{table}

\noindent 
The central problem in the Higgs field is the perturbation study after the spontaneous symmetry breaking. 
The inverted oscillator in QM does not have spontaneous symmetry breaking because the tunneling between different vacuums can happen, but it meets the same perturbation problem as in QFT.
The spontaneous symmetry breaking is necessary to introduce for avoiding the unstable vacuum for the negative mass squared case.
The unperturbed Hamiltonian cannot be the inverted oscillator as in QFT. 
The issue is avoided in the adaptive perturbation method because the interacting terms appear in the unperturbed Hamiltonian. 
Due to the successful application of the adaptive perturbation method to QM, QFT should be applied well.   
   
\section{Outlook}
\label{sec:4}
\noindent
The accurate study of $\langle x^2\rangle$ demonstrated that the adaptive perturbation method \cite{Weinstein:2005kw, Weinstein:2005kx} can study correlation functions. 
Confinement and asymptotic freedom are open questions in theoretical physics. 
Therefore, it is necessary to compute the correlation functions for all values of parameters. 
In general, the strongly coupled QFT only can be studied by the lattice method. 
It is hard to know whether the lattice method is correct due to the continuum limit. 
The adaptive perturbation method can be applied to lattice QFT and provides more clues to the open questions. 
\\

\noindent
To interpret the origin of the mass, the mechanism of the spontaneous symmetry breaking is necessary at this moment. 
To explore the mystery, one can study the single scalar field theory. 
The spontaneous symmetry breaking occurs in the single scalar field theory with the negative mass squared term. 
The lattice scalar field cannot obtain the vacuum expectation value because the different vacuums can tunnel. 
Due to the difference, one can apply the adaptive perturbation method to the lattice and continuum scalar field theories. 
It should be possible to help explore the concepts of spontaneous symmetry breaking before the study of the standard model of particle physics.
\\

\noindent
In the standard model of particle physics, the excitation of the Higgs field produces the Higgs boson \cite{Sirunyan:2017tqd}. 
The Higgs boson is ill-defined when a coupling constant vanishes because the vacuum expectation value of the Higgs field is inversely proportional to the square root of a coupling constant.  
Hence the weak-coupling physics in the Higgs field possibly cannot use the weak-coupling perturbation to give complete information. 
Theoretical studies ignored the problem before. 
Although LHC confirmed the theoretical studies about the Higgs field, the ill-defined problem possibly appears after reducing the statistical error. 
Our accurate study should demonstrate the applicability for the inverted mass case and sheds light on the relevant problems of the Higgs field.

\section*{Acknowledgments}
\noindent
The author would like to thank Su-Kuan Chu for his useful discussion and thank Nan-Peng Ma for his encouragement.
\\

\noindent
The author was supported by the Post-Doctoral International Exchange Program; 
China Postdoctoral Science Foundation, Postdoctoral General Funding: Second Class (Grant No. 2019M652926); 
Science and Technology Program of Guangzhou (Grant No. 2019050001).



\baselineskip 22pt
\begin{thebibliography}{99}

\bibitem{Hioe:1978jj}
F.~T.~Hioe, D.~Macmillen and E.~W.~Montroll,
``Quantum Theory of Anharmonic Oscillators: Energy Levels of a Single and a Pair of Coupled Oscillators with Quartic Coupling,''
Phys. Rept. \textbf{43}, 305-335 (1978)
doi:10.1016/0370-1573(78)90097-2

\bibitem{Ma:2018efs} 
  C.~T.~Ma,
  ``Parity Anomaly and Duality Web,''
  Fortsch.\ Phys.\  {\bf 66}, no. 8-9, 1800045 (2018)
  doi:10.1002/prop.201800045
  [arXiv:1802.08959 [hep-th]].
  
\bibitem{Sirunyan:2017tqd}
A.~M.~Sirunyan \textit{et al.} [CMS],
``Constraints on anomalous Higgs boson couplings using production and decay information in the four-lepton final state,''
Phys. Lett. B \textbf{775}, 1-24 (2017)
doi:10.1016/j.physletb.2017.10.021
[arXiv:1707.00541 [hep-ex]].

\bibitem{Coleman:1980aw}
S.~R.~Coleman and F.~De Luccia,
``Gravitational Effects on and of Vacuum Decay,''
Phys. Rev. D \textbf{21}, 3305 (1980)
doi:10.1103/PhysRevD.21.3305

\bibitem{Stone:1975bd}
M.~Stone,
``The Lifetime and Decay of Excited Vacuum States of a Field Theory Associated with Nonabsolute Minima of Its Effective Potential,''
Phys. Rev. D \textbf{14}, 3568 (1976)
doi:10.1103/PhysRevD.14.3568

\bibitem{Frampton:1976kf}
P.~H.~Frampton,
``Vacuum Instability and Higgs Scalar Mass,''
Phys. Rev. Lett. \textbf{37}, 1378 (1976)
[erratum: Phys. Rev. Lett. \textbf{37}, 1716 (1976)]
doi:10.1103/PhysRevLett.37.1378

\bibitem{Frampton:1976pb}
P.~H.~Frampton,
``Consequences of Vacuum Instability in Quantum Field Theory,''
Phys. Rev. D \textbf{15}, 2922 (1977)
doi:10.1103/PhysRevD.15.2922

\bibitem{Simon:1973yz}
B.~Simon and R.~B.~Griffiths,
``The phi(2)-to-the-4 field theory as a classical ising model,''
Commun. Math. Phys. \textbf{33}, 145-164 (1973)
doi:10.1007/BF01645626

\bibitem{Bargmann:1962zz}
V.~Bargmann,
``On the Representations of the Rotation Group,''
Rev. Mod. Phys. \textbf{34}, 829-845 (1962)
doi:10.1103/RevModPhys.34.829
 
\bibitem{Weinstein:2005kw} 
  M.~Weinstein,
  ``Adaptive perturbation theory. I. Quantum mechanics,''
  hep-th/0510159.
  
\bibitem{Weinstein:2005kx}
M.~Weinstein,
``Adaptive perturbation theory: Quantum mechanics and field theory,''
Nucl. Phys. B Proc. Suppl. \textbf{161}, 238-247 (2006)
doi:10.1016/j.nuclphysbps.2006.08.059
[arXiv:hep-th/0510160 [hep-th]].
  
\bibitem{Ma:2019pxd} 
  C.~T.~Ma,
  ``Adaptive Perturbation Method in Bosonic Quantum Mechanics,''
  arXiv:1911.08211 [quant-ph].
  
\bibitem{Curcio:2018} 
  F.~Curcio,
  ``Metodi di approssimazione delle energie di sistemi unidimensionali con potenziali polinomiali,''
  Tesi di Laurea Triennale (2017).
  
\bibitem{Ma:2020ipi}
C.~T.~Ma,
``Second-Order Perturbation in Adaptive Perturbation Method,''
[arXiv:2004.00842 [hep-th]].
  
  
  
\end{thebibliography}
\end{document}